# Transforming Science Learning Materials in the Era of Artificial Intelligence

Xiaoming Zhai[1], Kent Crippen[2]

*[1]AI4STEM Education Center, University of Georgia*

*[2]School of Teaching and Learning, University of Florida*

Abstract. The integration of artificial intelligence (AI) into science education is transforming the design and function of learning materials, offering new affordances for personalization, authenticity, and accessibility. This chapter examines how AI technologies are transforming science learning materials across six interrelated domains: 1) integrating AI into scientific practice, 2) enabling adaptive and personalized instruction, 3) facilitating interactive simulations, 4) generating multimodal content, 5) enhancing accessibility for diverse learners, and 6) promoting co-creation through AI-supported content development. These advancements enable learning materials to more accurately reflect contemporary scientific practice, catering to the diverse needs of students. For instance, AI support can enable students to engage in dynamic simulations, interact with real-time data, and explore science concepts through multimodal representations. Educators are increasingly collaborating with generative AI tools to develop timely and culturally responsive instructional resources. However, these innovations also raise critical ethical and pedagogical concerns, including issues of algorithmic bias, data privacy, transparency, and the need for human oversight. To ensure equitable and meaningful science learning, we emphasize the importance of designing AI-supported materials with careful attention to scientific integrity, inclusivity, and student agency. This chapter advocates for a responsible, ethical, and reflective approach to leveraging AI in science education, framing it as a catalyst for innovation while upholding core educational values.

## 1. Introduction

The integration of artificial intelligence (AI) into science education is profoundly reshaping the landscape of science learning materials. AI technologies are increasingly embedded in new forms of learning materials via various educational approaches—adaptive platforms (Tan et al., 2025), intelligent tutoring systems (Létourneau et al., 2025), multimodal simulations (Almasri, 2022), and generative tools (El Fathi et al., 2025)—fundamentally altering how students' access, process, and engage with scientific knowledge. These transformed learning materials not only support personalized learning trajectories but also reflect and simulate authentic scientific practice, that includes data modeling, hypothesis testing, and evidence-based reasoning (Erduran & Levrini, 2024; Herdiska & Zhai, 2024; Xu & Ouyang, 2022). In parallel with broader societal and scientific transformations, the integration of AI in learning materials (i.e., AI support) signals a pedagogical shift: moving beyond the static delivery of facts toward environments where complex data sets can be explored, hypotheses can be generated and tested in real-time, and patterns can be interpreted with AI-supported modeling tools. Not surprisingly, science learning materials have evolved in tandem to reflect these new affordances and to prepare students to work in AI-supported scientific contexts (Zhai & Krajcik, 2025).

Conventional textbooks and instructional designs frequently neglect the dynamic, context-sensitive, and multimodal characteristics that can be significantly enhanced by AI. They present knowledge as fixed, overlook the influence of learners' diverse perspectives, and offer limited interactivity that cannot adapt to students' questions or reasoning. These limitations restrict opportunities for authentic inquiry and evidence-based practice, areas where AI-supported materials can offer significant improvement. As scientific knowledge and practice become increasingly data-intensive and collaborative (Dong et al., 2017; Tenopir et al., 2011), there is a pressing need to redesign science curricula and materials to reflect these developments. Modern scientific inquiry frequently involves AI-supported tools for data collection, analysis, and modeling, techniques now central to disciplines ranging from climate science to genomics (Gil et al., 2014). For instance, in climate science, deep learning has been applied to predict extreme weather events and enhance climate model predictions by analyzing massive datasets from satellites and sensors (Reichstein et al., 2019). Similarly, in genomics, AI has accelerated the identification of gene–disease associations and advanced personalized medicine by enabling high-throughput analysis of sequencing data and predictive modeling of protein structures (Eraslan et al., 2019). Together, these cases illustrate how AI-supported practice has reshaped not only the scale of scientific data analysis but also the collaborative infrastructures that support discovery across domains.

To remain authentic and relevant, science learning materials must mirror these shifts by embedding AI-supported methods into instructional tasks and representations. Doing so can help students understand the nature of contemporary science, which is increasingly computational, algorithmic, and reliant on AI collaboration (Kitano, 2016). Without integrating these methods, science education risks presenting an outdated and incomplete view of how science is practiced. Moreover, students' roles are changing—they are not merely recipients of knowledge but active participants and co-constructors of meaning alongside intelligent systems, engaging in iterative processes of knowledge construction supported by AI tools that model, prompt, and adapt to learners' responses (Chen, 2025). This calls for a reimagining of how science is taught and learned, emphasizing flexibility, real-time responsiveness, and alignment of learning materials with the epistemic nature of contemporary science.

Moreover, AI is playing an increasingly important role in developing and designing learning materials with curriculum experts. Generative AI tools, such as ChatGPT and Claude, are being used collaboratively by educators to co-develop lesson plans, design formative assessments, and generate contextualized instructional content across subject areas and grade levels (Zhai, 2023). This partnership enhances the efficiency and scalability of curriculum development, allowing instructional designers to produce tailored materials that are adaptive to diverse learner needs. In addition, AI can rapidly analyze curriculum standards, student performance data, and scientific literature to recommend content sequences or highlight emerging concepts relevant to instruction (Su & Zhong, 2022). Such integration signifies a paradigmatic shift in the instructional design process, where human-AI co-authorship enables rapid prototyping and iterative refinement of educational materials. This collaborative process not only augments human creativity and productivity but also ensures a more responsive curriculum that can evolve with scientific advancements and societal needs. By leveraging large-scale datasets and natural language generation, AI enables educators to design tasks that are more closely aligned with real-world scientific inquiry (Crippen & Archambault, 2012), thereby fostering students' critical thinking and problem-solving skills.

Despite its promise, the integration of AI into science learning materials raises critical ethical and pedagogical concerns that must be urgently addressed. Unresolved issues, including algorithmic bias, misinformation, lack of transparency, and data privacy risks, pose significant threats to equitable and inclusive science learning (Akgun & Greenhow, 2022; Latif & Zhai, 2025). These risks are particularly pronounced when AI systems are used to create or curate content without adequate human oversight or validation. Without clear standards or safeguards, there is a danger of perpetuating existing educational inequities or misrepresenting scientific knowledge, especially for marginalized communities. Moreover, the opaqueness of many AI models complicates teachers' and learners' ability to critically evaluate AI-generated information, potentially undermining student and teacher agency in the classroom. These concerns highlight a pressing gap in the current educational landscape: the need to systematically examine and establish responsible and ethical norms, frameworks, and practices that ensure AI technologies support the transformation of science learning materials rather than compromise the goals of science education. The absence of such considerations could result in science learning materials that are technologically advanced but ethically fragile and pedagogically misaligned.

Given these gaps, this chapter examines the transformative potential of AI integration in science learning materials, articulating the changes brought by AI—both the opportunities and challenges associated with its integration in science learning materials. Drawing on current research, design examples, and educational frameworks, we examine key domains where AI is reshaping science learning materials, including embedded scientific practice (Bewersdorff et al., 2025), personalized learning (Yılmaz, 2024), interactive simulations (Erümit & Sarıalioğlu, 2025), multimodal content (Almasri 2022), accessibility (Lee et al., 2025), and content creation (Bewersdorff et al., 2025). While these domains can be considered separately, they are deeply interrelated: for instance, personalized learning depends on multimodal content and accessible design to be truly adaptive, while interactive simulations and embedded scientific practice together foster inquiry-rich experiences that also generate new opportunities for content creation. In practice, advances in one domain often catalyse developments in another, forming a network of mutually reinforcing transformations that together redefine what it means to design and use science learning materials. We engage with the responsible (how) and ethical (why) dimensions of these transformations, proposing solutions to guide the use of AI. By offering an analytical and normative lens, this chapter aims to support educators, developers, and decision-makers in designing inclusive, adaptive, and ethically informed science learning materials for the AI era.

## 2. Changes Brought by AI

### 2.1. AI embedded in scientific practice

Contemporary science increasingly relies on AI for core practice, such as large-scale data analysis, predictive modeling, and hypothesis generation (see Chapter 2 of this book). These transformations reflect fundamental changes in the epistemic foundations of scientific research, where AI is not merely a tool but an active assistant or collaborator in scientific discovery (Bianchini et al., 2022). As scientists integrate AI in fields like climate science and genomics, where complex simulations and pattern recognition are essential (Kitano, 2016), it becomes imperative that science education reflects these changes. Science learning materials must evolve to reflect such authentic practice, enabling students to develop an understanding of how modern science is conducted and how AI is reshaping its methods, goals, and ethical considerations (Vartiainen et al., 2021).

A salient example of this shift can be observed in the case of AlphaFold, an AI system developed by DeepMind to predict protein structures with remarkable accuracy (Jumper et al., 2021). AlphaFold employs deep learning algorithms, specifically a neural network architecture trained on extensive protein sequence and structure databases, to predict three-dimensional protein configurations from amino acid sequences. Its methodology integrates multiple sequence alignments and attention mechanisms to infer spatial relationships among residues, enabling predictions that rival the accuracy of experiments. This tool has transformed genomics and structural biology by resolving the structures of proteins that were previously difficult or impossible to characterize using traditional techniques such as X-ray crystallography or cryo-electron microscopy. As a result, AlphaFold has been integrated into research workflows to accelerate drug discovery, understand genetic diseases, and interpret evolutionary relationships among organisms. In response, educational initiatives have begun integrating AlphaFold-inspired activities into biology curricula (Samson, 2025). These activities include AI-supported simulations where students explore protein folding, test hypotheses, and interpret structural data using tools similar to those employed by professional scientists.

In science classrooms, researchers have explored the integration of AI in scientific inquiry activities. For example, embedded AI tools such as Google's Teachable Machine can be used to guide students in training the algorithm to identify specific features of plants, thus mimicking scientists' work (see Figure 1 for an example). Working in small groups, students first collect and label their own sets of plant images to serve as training data. They consider which features, such as leaf shape, flower color, or stem structure, would best distinguish one plant type from another, mirroring how scientists identify and operationalize variables. They upload their datasets, organizing them into labeled classes, and training the algorithm to recognize patterns across their images. Students then test the algorithm with new images, compare its classifications with their own judgments, and refine the datasets and labels to improve accuracy. Through this process, they engage in core scientific practice, including the design of investigations, analyzing and interpreting data, constructing and revising models, and validating findings against empirical evidence. Such instructional experiences provide students with a window into real-world scientific processes enhanced by AI, thereby bridging the gap between classroom learning and contemporary scientific practice. In addition to Google's Teachable Machine, teacher-friendly applications such as Microsoft Lobe (https://www.lobe.ai), IBM Watson Studio (https://dataplatform.cloud.ibm.com), MIT App Inventor (https://appinventor.mit.edu), and Machine Learning for Kids (https://machinelearningforkids.co.uk) are also available to achieve this goal.

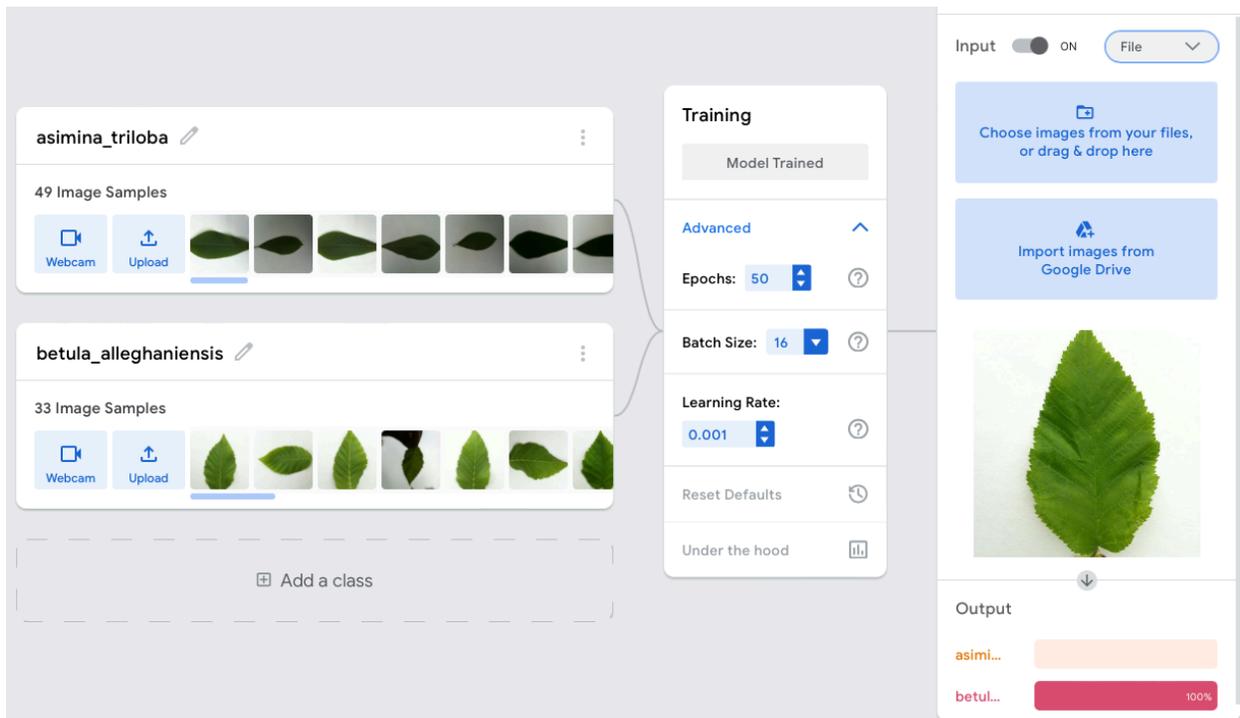

Figure 1. An example of students using Google Teachable Machine to train a model to differentiate two types of leaves.

The Google Teachable Machine case exemplifies the potential for reimagining science learning materials, considering contemporary scientific practice, which increasingly relies on AI. Traditional science learning materials, often presented as linear, static, and text-driven, consistently underperform when compared to dynamic, interactive alternatives. For example, students using simulations or sequenced frames outperformed those relying on conventional static diagrams in kinematics study (Mešić et al., 2015). Moreover, frameworks like Duschl's Evidence-Explanation Continuum make explicit how scientific reasoning evolves through iterative, data-rich cycles of measurement, modeling, and explanation, cycles that static texts rarely support (Duschl et al., 2021). As exemplified by tools like AlphaFold, AI enables scientists to make discoveries that depend on large-scale data analysis, complex simulations, and predictive modeling—capabilities that are beyond the scope of traditional science instruction. Yet, most current learning materials do not provide students with opportunities to engage in AI-supported practice. Without meaningful updates to reflect how science is now practiced, students are left with an outdated view of inquiry and discovery, missing key epistemic insights into how AI is transforming the nature of scientific knowledge production and application.

## 2.2. Adaptive and personalized learning experiences

The advent of AI presents an unprecedented opportunity to create new forms of science learning materials specifically designed to support adaptive (i.e., real-time, responsive adjustments to pace and content) and personalized learning (i.e., a broad combination of technology, including adaptive options, with pedagogy, curriculum, and student agency). Traditional learning materials, such as textbooks and static online modules, are often not inclusive or responsive because their fixed, text-driven structures limit representation and adaptability (Smith et al., 2022). They reproduce narrow views of scientific knowledge, omitting diverse perspectives, and are less effective than dynamic, multimodal resources in supporting varied reasoning approaches (Stelzer et al., 2009). Moreover, research on adaptive assessments demonstrates how static formats fail to accommodate learners' prior knowledge, underscoring the need for more responsive designs (Zhai et al., 2021).

AI-supported adaptive learning systems can dynamically generate and adjust learning content based on real-time data collected from student interactions (Kabudi et al., 2021; Létourneau et al. 2025; Wang et al. 2024). These systems analyze learner inputs, such as quiz responses, engagement metrics, and behavioral data, to develop detailed learner profiles that guide individualized instruction. Through this process, AI can deliver learning materials that adjust to each student's unique pace, scaffolding, modality, and complexity, aligning with their individual progression. Such learning environments are not only more responsive but also more inclusive, particularly for students from linguistically and culturally diverse backgrounds or those with learning differences (Nyaaba et al., 2024).

Several emerging applications of AI demonstrate the possibilities of creating novel forms of science learning materials, particularly for adaptive and personalized learning. In biology, systems such as the Learning Intelligent Tutoring System (LITS) offer interactive tutorials on cellular functions that adjust the depth and style of explanation in real-time. For example, Schmucker et al. (2024) introduced a LLM-based conversational tutoring system, which provides adaptive support to learners. The LITS system automatically generates a tutoring script from lesson texts using LLMs, which specify review questions, solutions, and learning expectations in a format that can be easily edited by instructors. The system is orchestrated through two conversational agents—Ruffle, a student, and Riley, a professor—who guide the learner in a learning-by-teaching scenario. Ruffle prompts the learner to explain concepts, while Riley provides scaffolding, corrective feedback, and clarification when misconceptions are detected. This dual-agent design enables adaptive dialogues that follow the inner- and outer-loop structures typical of intelligent tutoring systems, but with significantly reduced authoring costs compared to traditional CTSs. Empirical evaluations with biology lessons on cell organelles demonstrated that learners rated the system as more engaging, coherent, and helpful than simpler QA chatbots, and that many completed the entire conversational workflow. Although no significant differences in short-term test performance were observed compared to reading conditions, interaction analyses revealed that learners who engaged more deeply in conversations achieved higher learning gains, highlighting the potential of LLM-driven systems to foster adaptive, personalized science learning at scale.

The shift toward adaptive AI-supported learning materials represents more than a technological innovation—it demands a pedagogical transformation. Such materials can offer differentiated entry points into scientific content, enabling all students to engage meaningfully with core ideas and practice. Unlike conventional resources, which often assume a uniform baseline of understanding, AI-supported materials can be fluid, multimodal, and responsive to formative data. They have the potential to foster not only content mastery but also learner autonomy, self-regulation, and metacognitive reflection. To realize this potential, educators and developers must collaborate to design materials that utilize AI not only to customize learning paths but also to deepen scientific inquiry, simulate authentic practice, and promote educational accessibility. In doing so, the development of new AI-supported learning materials becomes essential—not just for personalization, but for transforming science education into a more adaptive, responsive, and student-centered enterprise.

### 2.3. Contextualized and interactive simulations

AI enables the development of new forms of science learning materials by powering contextualized and interactive simulations that reflect real-world phenomena (Almasri, 2022) Going far beyond traditional, static digital labs, these AI-supported simulations may dynamically adjust parameters, prompts, and investigative pathways based on students' real-time input, inquiry strategies, and learning context (Adair et al., 2023; Ben-Zion et al., 2025). This shift transforms students' roles and experience with simulations, from passive users to creators and designers. This new form of AI-supported simulations also enables highly personalized, scenario-based environments that can replicate the complexity and iterative nature of modern scientific research.

Figure 2 shows an example of transforming a sketch into interactive simulations when students learn projectile motion. In this example, a student drew a symbolic ball and wall with variables such as speed, gravitational coefficient, and the distance from the ball to the wall, and then supplied the sketch to ChatGPT (Figure 2, left). ChatGPT then generates the code for the HTML file, which is opened in a browser. After opening the HTML file, students can study projectile motion using the simulation by manipulating variables such as the speed of release, running the simulation, and observing the resulting path of the ball as it interacts with the wall under the force of gravity (Figure 2, right). In this example, AI simulations not only enhance engagement but also serve as personalized learning materials that involve students in creating, using, and testing simulations of observable phenomena for science learning. However, it also raises important questions about the accuracy of the underlying simulation. Thoughtful pedagogy such as engaged critiquing is needed to leverage these student-AI generated simulations for epistemic purposes.

| Student Drawing | Ball Motion Simulation |
|---|---|
| 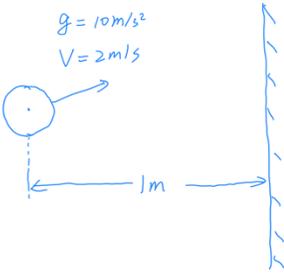 Prompt: Create a simulation to show the motion of the ball before and after hitting the wall according to the information on the drawing | 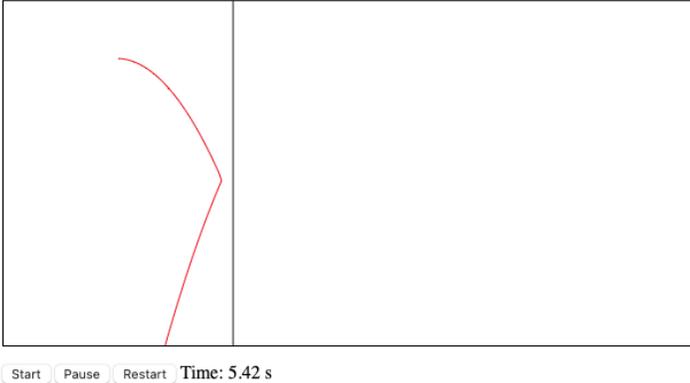 |

| attached. The simulation should include start, pause, and restart buttons, a timer. Once start, it should show the trajectory of the ball. | |

Figure 2. Sketch to simulation. Students use ChatGPT to create an interactive simulation based on a hand sketch[1].

Besides the simple on-the-fly simulations, AI can generate and model complex scientific phenomena that are otherwise inaccessible without significant efforts in coding and testing. For example, Ben-Zion et al. (2024) discuss leveraging generative AI to rapidly develop physics simulations, allowing educators to create personalized virtual labs such as a pendulum simulation (see Figure 3). Unlike PhET simulations (Finkelstein, 2006), which are professionally developed, thoroughly validated, and designed for broad classroom use, such AI-generated simulations are highly customizable, allowing educators and even students to generate and refine simulations tailored to specific concepts or edge cases not covered by existing tools. While PhET offers stable, classroom-ready models, AI simulations uniquely position learners as co-designers, putting them in charge of the iterative processes of creation, validation, and refinement, as the practice of science. Moreover, students can explore complex physical phenomena, such as quantum mechanics and wave functions, in a controlled, risk-free environment, thereby overcoming traditional barriers related to safety and resource limitations (Ben-Zion et al., 2024). In biology, similar simulations can model disease transmission with adjustable mutation rates and vaccine efficacy, enabling students to explore causal reasoning and systems thinking. These simulations are not merely interactive tools; they are intelligent learning materials that evolve in response to student inputs, fostering a deeper understanding through contextualized, personalized scenarios.

---

[1] This example idea credited to Jennifer Kleim, a Ph.D. student of the first author.

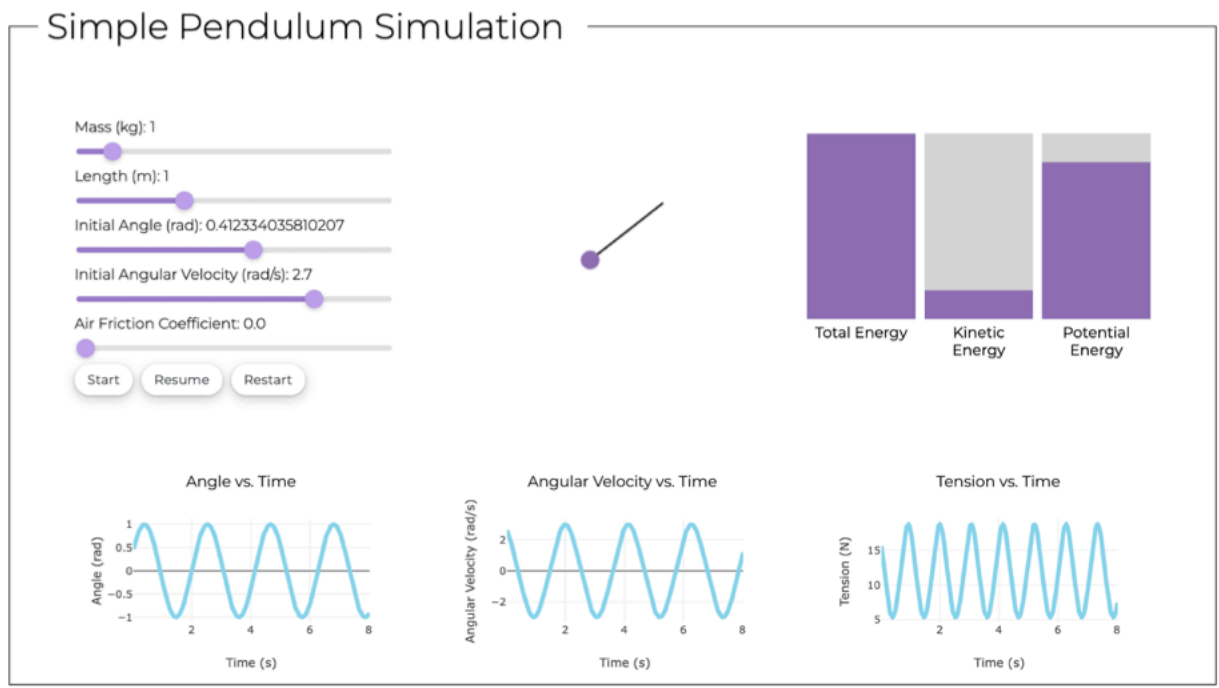

Figure 3. Pendulum simulation featuring large-angle dynamics, friction effects, pendulum tension graph, and energy distribution visualization. Reprint from Ben-Zion et al. (2024), arXiv:2412.07482 under CC BY 4.0 (https://creativecommons.org/licenses/by/4.0/). No changes were made.

Moreover, AI simulations also introduce new material formats through the inclusion of adaptive avatars and conversational agents. These elements support collaborative inquiry by facilitating dialogue, offering counterarguments, and prompting metacognitive reflection (Makransky et al., 2019). A recent review study suggests that AI-supported virtual agents can engage students in decision-making tasks that mirror authentic challenges, yielding medium overall effects on learning (g = 0.43) when compared to simulations without AI-supported agents (Dai et al., 2024). In science classrooms, AI-supported virtual agents function as adaptive guides, providing feedback, hints, and personalized support to help students navigate complex inquiry tasks and decision-making processes (Thompson & McGill, 2017; Ward et al., 2013). As such, these new forms of learning materials embed students within scientific discourse communities, encouraging not only content mastery but also the development of communication and argumentation skills essential to science literacy.

To capitalize on these advancements, the design of AI-supported simulations must be rooted in robust pedagogical frameworks. Educators must ensure that these materials go beyond delivering automated answers to cultivating critical inquiry, reflection, and agency. Especially given that hallucinations represent a persistent issue impacting the validity and reliability of any output from AI (Ji et al., 2022). This can be partially achieved by using inquiry-based and modeling practices to structure the use of simulations. Rather than positioning AI as a source of solutions, teachers can frame it as a partner in investigation, prompting students to evaluate, refine, and extend its outputs. Embedding opportunities for reflection, such as asking students to explain why a simulation behaved in a certain way or to critique the validity of AI-generated results, encourages deeper reasoning. Agency can be further supported by allowing students to choose parameters, datasets, or approaches, positioning them as decision-makers in the learning process. In this way, AI becomes a catalyst for evidence-based thinking and metacognitive growth, rather than a shortcut to answers. When properly integrated, AI-supported simulations represent a significant evolution in science learning materials: from passive, uniform resources to dynamic, personalized, and context-sensitive environments that prepare students for authentic scientific engagement.

## 2.4. Integration of multimodal and dynamic content

AI supports automatic generation and analysis of multimodal materials (e.g., text, image, video, data), leading to a fundamental transformation of science learning materials. Traditionally limited to static, text-centric formats, science materials are now evolving into dynamic, interactive, and richly multimodal resources (Stelzer et al., 2009; Zhai et al., 2021). AI can automatically generate visualizations from raw data, produce narrated video summaries of complex experiments, or create interactive diagrams that adapt in real time to student interactions (Bewersdorff et al., 2025). These affordances enable science content to be represented and for those representations to change dynamically in ways that are more aligned with how knowledge is constructed and communicated in professional scientific practice (Lee et al., 2023). As a result, learners benefit from more engaging and accessible materials that support diverse preferences, learning profiles, and disciplinary literacy skills.

Generative AI can also be used to create visualized learning materials, such as analogies to science concepts, to facilitate learning. For example, electric voltage and current have been challenging for middle school students to learn due to their abstract nature. To help students understand the concepts, Cao et al. (2023) used generative AI to generate multimodal learning materials, images, and videos of water pressure/current that analogize electric voltage and current, reducing the learning difficulty. Chemistry concepts, such as electrolysis and gas emission, can be difficult for students to visualize because they involve abstract or microscopic processes. To address this, Akaygun and Kilic (2025) had preservice teachers use generative AI to create text- and image-based visual analogies. The AI-generated materials were perceived as attractive and engaging, facilitating students' understanding of the macroscopic aspects of chemical processes. However, teachers also noted limitations in scientific plausibility and tool constraints, underscoring the need for teacher guidance in their use.

Another major shift enabled by AI is the capacity to generate up-to-date, contextually relevant content that reflects the rapidly evolving nature of scientific knowledge. Science learning content needs to reflect the contemporary and current development of science, but curating and transforming these new developments into accessible learning materials has traditionally been infeasible. With the development of AI, tools such as Elicit and Semantic Scholar can now scan and synthesize emerging research, providing real-time summaries and curated insights that can be directly integrated into classroom science learning materials. These systems can suggest research questions, generate comparisons, summaries, or explanations. Notebook LM, a current product from Google, allows students to upload class materials, such as lab guides, readings, published papers, or data sets, and then ask questions, receive summaries, or generate explanations grounded in those sources. Its purpose is to help students connect ideas across documents, making complex science content more accessible while keeping inquiry focused on teacher-curated materials. This interaction positions students as investigators, using AI to highlight connections, identify gaps in their understanding, and test their interpretations against multiple sources, including primary, peer-reviewed, empirical research publications. Teachers could further scaffold reflection by asking students to critique the AI's outputs, reinforcing skills in evidence evaluation and scientific reasoning while preventing over-reliance on automated responses. This type of use enables science learning resources to remain current and epistemologically authentic, exposing students to the provisional and dynamic character of scientific understanding. Using these AI-supported science learning materials, students can enhance their awareness of the evolving nature of science, highlighting how such materials support deeper epistemic cognition.

However, the shift toward AI-supported materials demands critical attention to their educational validity and use. There is an inherent risk of over-reliance on algorithmic outputs, which may oversimplify complex scientific concepts, present the concepts inaccurately, or unintentionally perpetuate bias. Therefore, the transformation of science learning materials must be guided by intentional design, with educators equipped to interpret, adapt, and contextualize AI-generated content. As AI continues to reshape how learning materials are created and used, it is essential that its role be understood as augmenting—and not replacing—the pedagogical expertise of teachers and the epistemic engagement of students (Wang et al., 2025).

### 2.5. Language and accessibility support

AI-supported translation, summarization, and speech-to-text features are transforming the accessibility of science learning materials, particularly for multilingual and neurodiverse learners. These tools enable the real-time adaptation of complex scientific texts into multiple languages or simplified summaries, making the content more accessible to students with varied linguistic backgrounds and reading abilities. For example, platforms such as Microsoft's Reading Progress were integrated into digital learning environments to promote English Language Learners' (ELLs') reading aloud, thereby enhancing access to learning materials for ELLs and students with auditory processing challenges (Jose, 2025).

Moreover, AI technologies support neurodiverse learners by facilitating multimodal access to information (Panjwani-Charania & Zhai, 2024). Speech-to-text tools can convert students' spoken responses into written form, supporting those with dyslexia, dysgraphia, or fine motor difficulties. Likewise, summarization algorithms help reduce cognitive load by distilling key concepts and eliminating extraneous information from scientific texts. A study by Thaqi et al. (2024) using an AI-supported reading assistant to help learners, particularly with lower learning proficiency or dyslexia, address challenges such as unfamiliar vocabulary and complex sentence structures by providing contextual explanations, rephrased content, and multilingual support. This approach highlights AI's potential to enhance the reading experience and support science learning for students with lower reading proficiency or neurodiverse learners.

AI can also be used to revise text, adaptable to students' level of language proficiency. For example, the study by Bewersdorff et al. (2025) illustrates how students attempt to understand the function of an insect's compound eye using an image sourced from Wikipedia. To support students at both grades 5 and 12, they leveraged generative AI by enabling adaptive interaction with visual materials and enriching the accompanying textual content, taking into account students' prior knowledge and background. Figure 4 illustrates how generative AI analyzes the image and delivers information tailored to learners' levels of competence.

Figure 4. Explanatory information generated by GPT-4V for 5th and 12th-grade students regarding a graphical representation of an insect's eyes. Reprinted from Bewersdorff, A., Hartmann, C., Hornberger, M., Seßler, K., Bannert, M., Kasneci, E., Kasneci, G., Zhai, X., & Nerdel, C. (2024). Taking the next step with generative artificial intelligence: The transformative role of multimodal large language models in science education. Learning and Individual Differences, 118, 102601. https://doi.org/10.1016/j.lindif.2024.102601 — licensed under Creative Commons Attribution 4.0 (CC BY 4.0) (https://creativecommons.org/licenses/by/4.0/). No changes were made.

In addition, AI can also be used to promote inclusive participation by allowing learners to interact with materials in the modalities that best suit their cognitive strengths and preferences. For instance, a neurodiverse student may prefer auditory explanations over written texts, while a multilingual student might benefit from side-by-side translations. When AI tools are integrated into learning platforms, they create opportunities for personalized and dignified engagement with science content, reflecting a more equitable vision of education.

However, the deployment of AI for accessibility must be underpinned by careful design considerations. Developers and educators must ensure that automated translations accurately convey the meaning of scientific terminology and that summarization tools preserve critical disciplinary knowledge. This requires aligning language models with domain-specific vocabularies and validation processes to check the fidelity of outputs, ensuring that nuanced concepts such as "force" or "energy" are not oversimplified or mistranslated. Yet, accomplishing this in real-time is difficult, as it demands continuous monitoring, expert oversight, and safeguards that extend beyond what individual teachers can realistically manage, which underscores the shared responsibility of designers, institutions, and policymakers in supporting responsible and ethical AI use in science education (see chapter 3 in this book). For example, ethical concerns regarding data privacy, algorithmic fairness, and cultural sensitivity must be addressed to ensure these tools genuinely serve the needs of diverse learners. In this way, AI can become a powerful ally in the pursuit of universally designed science learning materials that respect and uplift all students.

## 2.6. AI as a co-author

Educators and curriculum developers are increasingly turning to generative AI tools to co-create lesson plans, assessments, simulations, and activities, marking a significant shift in the development of instructional materials. This shift is not merely a matter of efficiency—it represents a transformation in the role of teachers and developers as co-designers working alongside intelligent systems. Tools like ChatGPT, Claude, and Copilot are enabling educators to rapidly generate diverse instructional materials tailored to varying cognitive levels and subject domains. For instance, Lee and Zhai (2024) engaged pre-service science teachers in using GPT-based systems to produce lesson plans for science teaching, allowing educators to quickly iterate and customize learning materials based on students' needs. Pre-service teachers demonstrated moderately high competence in integrating ChatGPT into lesson planning, with particularly strong alignment to instructional strategies and curriculum goals. They envisioned diverse applications across various scientific domains, utilizing ChatGPT for questioning, individualized support, and formative assessment. However, they struggled with selecting appropriate functions, sometimes misapplied the tool (e.g., relying on hallucinated resources), and raised concerns about misinformation and student overdependence. These findings highlight both the promise of AI integration and the need for teacher guidance and systemic safeguards.

One of the most profound changes brought by AI co-authorship is the ability to produce contextually relevant materials on demand. This responsiveness allows educational content to remain current with scientific advancements and sociocultural shifts. For example, Nyaaba et al. (in press) introduced the *Culturally Responsive Lesson Planner (CRLP)*, a theory-grounded AI tool designed to embed cultural and linguistic context into science lesson plans. Built upon Culturally Responsive Pedagogy (CRP), Funds of Knowledge, and the GenAI-Culturally Responsive Science Assessment (GenAI-CRSciA) framework (Nyaaba et al., 2024), the CRLP employs an Interactive Semi-Automated (ISA) prompting system that guides teachers through a step-by-step dialogue to supply essential cultural parameters such as local language, indigenous knowledge, and community practices. Rather than producing generic outputs, the system requires teachers to provide contextual details that the AI then weaves throughout objectives, activities, assessments, and extension tasks. When tested in Ghana's Ashanti Region, the CRLP generated Grade 7 science lessons on states of matter that integrated Asante Twi terminology (e.g., *ɛpono* for wooden furniture, *dadeɛ nkuto* for metal objects, *aboɔ* for stones) and locally relevant cultural practices like fufu preparation and palm oil processing. Comparative evaluations by expert reviewers showed that CRLP-produced lessons contained more than double the cultural elements of standard GPT-4o outputs (48.5 vs. 21) and achieved higher ratings in accuracy (1.8 vs. 1.2) and curriculum relevance (2.0 vs. 1.3). Importantly, the CRLP also included contextualized teaching resources, bilingual vocabulary, and culturally responsive assessments, while inviting teacher input and modification before finalizing the lesson plan. Beyond improving cultural authenticity, the iterative dialogue with the CRLP fostered teacher agency and AI literacy, positioning GenAI not as a static generator but as a collaborative co-author that adapts content to local contexts in real time. Taken together, the CRLP exemplifies how AI co-authorship can mitigate Western-centric biases and enable the rapid creation of educational materials that are not only scientifically accurate but also culturally meaningful across diverse global contexts.

Generative AI can also democratize the design process, enabling educators with limited time or technical resources to participate in the creation of high-quality content. Studies by Kim and Wargo (2025) show that teachers in rural or underfunded schools reported improved confidence and perceived usefulness when using AI tools to generate science learning activities. This shift reduces reliance on pre-packaged curriculum and empowers teachers to act as adaptive, responsive designers of learning experiences. Moreover, AI systems can provide scaffolding during content creation by suggesting revisions, checking factual accuracy, or enhancing clarity, effectively serving as collaborative writing partners.

However, the integration of generative AI into curriculum development also raises critical questions about authorship, quality assurance, and pedagogical integrity. It is essential that AI-supported materials undergo rigorous review by educators to ensure alignment with learning objectives, standards, and ethical norms. While this requirement creates a tension with the benefits of real-time adaptivity described previously, it also highlights a productive balance: rapid responsiveness must be coupled with deliberate oversight to ensure that adaptivity enhances, rather than undermines, responsible and ethical practice. Without careful oversight, there is a risk of reinforcing biases, inaccuracies, or superficial content. As such, professional development and institutional guidelines are needed to support educators in leveraging generative AI responsibly and effectively. When appropriately integrated, AI co-authorship holds the potential to enhance creativity, efficiency, and accessibility in science education.

## 3. Responsible and Ethical Uses of AI in Response to the Changes

While the integration of AI into the development and design of science learning materials offers unprecedented opportunities, it is equally important to embed responsible and ethical principles transparently into both their construction and deployment (as articulated in the REP framework, Chapter 3 of this book). Such efforts are essential to ensure that AI-supported educational innovations not only enhance learning outcomes but also promote accessibility, inclusion, and human-centered values in science education. The thoughtful implementation of AI should be guided by frameworks that prioritize student empowerment, educator agency, and social justice. In the following sections, we elaborate on the key dimensions of ethical considerations that must shape the responsible use of science learning materials in light of the changes brought by AI.

### 3.1. Promote Transparency and Critical AI Literacy in Scientific Practice

As students increasingly encounter AI while engaging in scientific practice—such as creating models, analyzing large datasets, conducting virtual experiments, and generating hypotheses—it becomes crucial to ensure that these experiences are grounded in *transparency and explainability*. Students must be provided with opportunities to build foundational AI literacy that encompasses not only a technical understanding of how AI algorithms operate, but also a reflective awareness of how training data, model architectures, and design assumptions influence AI outputs, potentially introducing epistemic bias, privileging certain forms of knowledge or perspectives. *Transparency* and *human oversight* are necessary. AI systems used in scientific contexts should be accompanied by documentation and user-facing explanations that enable students to critically evaluate the sources, processes, and reliability of AI-generated content, yet recent research demonstrates that even many applied researchers disagree on or misunderstand explainability outputs—calling for careful design and scaffolding rather than assuming explanations mean understanding (Krishna et al., 2023). For example, Zhai, Shi, and Nehm (2021) show that even in machine learning–based science assessments, human oversight is essential because students and educators must interpret outputs critically and reconcile them with disciplinary reasoning.

Educators must scaffold learning experiences that encourage students to critique AI outputs, identify errors or inconsistencies, and make informed decisions about when and how to use AI tools. Additionally, institutions must ensure that the integration of AI in science learning aligns with ethical review standards, minimizes biases in representations and predictions, and upholds the integrity of scientific inquiry by treating AI as a tool to augment, not replace, human reasoning.

### 3.2. Design Accessible, Transparent, and Privacy-Conscious Personalized Learning

AI-supported personalization presents compelling opportunities to deliver content, pacing, and feedback tailored to learners' individual needs, skills, and interests. However, this personalization must be designed with *accessibility* at its core. Ethically adaptive systems should be sensitive to cultural, linguistic, and socioeconomic differences among learners, ensuring that recommendations or adaptations do not inadvertently perpetuate systemic inequities. For example, developers must audit AI models for bias in data collection and algorithmic decision-making to uphold *fairness*, while policymakers and educators must advocate for *transparency* in algorithmic design and outputs. Privacy and data protection are also critical, as adaptive systems often rely on the ongoing collection of sensitive behavioral and academic data. Clear consent mechanisms, data anonymization strategies, and compliance with data protection laws (e.g., FERPA, GDPR) must be enforced. Importantly, adaptive learning environments should maintain the agency of students and educators, allowing them to adjust or override AI-generated recommendations. Transparency reports, user dashboards, and opt-out features are crucial for promoting trust and digital autonomy. Research on intelligent tutoring and personalization underscores both the promise and risks of such systems: they can scaffold inquiry and self-directed learning, but they also risk amplifying inequities if not carefully designed and monitored (Holmes et al., 2019).

### 3.3. Ensure Scientific Integrity and Human Oversight in AI Simulations

AI has opened the door to richly contextualized and responsive simulations that enable students to engage with complex scientific phenomena in ways that were previously inaccessible due to safety, cost, or logistical constraints. For instance, students can now manipulate variables in a virtual chemical reaction, observe astrophysical events over time, or conduct experiments with AI-generated feedback. These tools hold the potential to foster deep conceptual understanding and active learning, but only if they are grounded in scientifically accurate models and designed with pedagogical intention. Developers must work closely with scientists, science educators, and curriculum designers to ensure that simulations accurately reflect representations and support key learning objectives. Educators should be provided with professional development to effectively integrate these tools, including strategies for guiding inquiry, diagnosing misconceptions, and encouraging students to reflect on the scientific reasoning behind observed outcomes. Moreover, students should be supported in understanding the limitations of simulations and the assumptions underlying the models they explore, promoting a nuanced and critical engagement with digital tools. This is evident when comparing robust, research-validated platforms, such as PhET (Finkelstein, 2006), with newly emerging AI-generated simulations (Ben-Zion et al., 2025), which can be highly flexible but require additional validation and oversight.

### 3.4. Curate Accurate and Pedagogically Coherent Multimodal Content

AI technologies now enable the generation and analysis of complex, multimodal learning content that includes not only written texts but also audio explanations, diagrams, infographics, interactive graphs, and videos. Such content has the potential to make abstract scientific concepts more tangible and engaging, particularly for students who benefit from visual or auditory learning modalities. However, this abundance of dynamic content must be curated with a focus on scientific accuracy, cognitive load, and instructional coherence. Developers should be held accountable for validating the sources, claims, and data underpinning AI-generated materials to promote transparency. Cross-disciplinary teams of content experts, educators, and technologists should collaborate to design and review AI-generated learning resources. Furthermore, multimodal systems should incorporate features that enable differentiation and Universal Design for Learning (UDL), ensuring that content is accessible, inclusive, and adaptable to diverse learner needs. Educators also require training and support to critically evaluate and effectively integrate multimodal AI content into their lesson planning and formative assessment practice. Studies have shown, for instance, that multimodal and interactive formats outperform static text in supporting conceptual understanding in physics (Stelzer et al., 2009) and that multimodal cues, such as facial expressions, can help diagnose and address student misconceptions in kinematics (Liaw et al., 2021).

### *3.5. Advance Inclusive Access Through Ethical Language and Accessibility Tools*

One of AI's most transformative potentials in attending to science learning materials lies in its ability to bridge linguistic and accessibility gaps. Tools that offer real-time language translation, text simplification, voice-to-text transcription, and multimodal summarization can democratize access to scientific knowledge for students who are multilingual, neurodivergent, or have disabilities. However, ethical implementation requires more than just technical inclusion. These tools should be developed in consultation with diverse user communities to ensure they are culturally responsive and affirming of students' linguistic identities (i.e., fairness). Rather than merely translating dominant forms of scientific discourse, AI systems should also validate and incorporate multiple ways of knowing and expressing scientific understanding. Interface design must emphasize usability across devices, connectivity levels, and contexts. Educators must also be equipped with pedagogical strategies for using these tools to foster agency and participation among all students. Accessibility, explainability, and transparency should be built into the DNA of AI learning platforms, not added as an afterthought. For example, Watters et al. (2021) demonstrated how AI-supported tools can improve access for visually impaired students in science classrooms, while Smith, Avraamidou, and Adams (2022) highlight the importance of culturally sustaining pedagogies to ensure that inclusive access extends beyond language or disability accommodations.

### 3.6. Maintain Human Oversight and Clarify AI's Role in Educational Co-Creation

The rise of generative AI introduces new opportunities for educators to co-create content such as assessment questions, lesson plans, explanatory texts, or even simulations. Yet, this co-authoring role demands a clear framework for accountability, transparency, and the ethical use of resources. Educators must remain the final arbiters of the quality, alignment, and appropriateness of AI-supported materials, ensuring that these resources meet educational goals and uphold professional standards and scientific integrity. Materials should clearly indicate the involvement of AI in their creation, allowing students and other stakeholders to engage with them critically and transparently. Moreover, the anthropomorphising of AI tools (i.e., describing or treating AI systems as if they possess human-like qualities)—in public discourse and classroom practice alike—must be actively countered. As Nash (2024) cautions, referring to AI as if it "knows," "thinks," or "believes" obscures its fundamental nature as a statistical inference tool. Educators should help students develop epistemic clarity, emphasizing that AI lacks consciousness, understanding, or intent. Teaching these distinctions fosters digital literacy and safeguards students' critical reasoning skills. When used responsibly, AI can support co-creation, ideation, and iteration, but it must always be contextualized within human values of autonomy, dignity, and inclusive participation. In this vision, AI becomes a collaborative instrument for enhancing inquiry and deepening engagement—a means of expanding, rather than narrowing, the role of human judgment in science education. This was illustrated in a study of pre-service teachers using ChatGPT for lesson planning, where participants generated creative science activities but also over-relied on AI outputs, underscoring the importance of human oversight (Herdiska & Zhai, 2024).

## 4. Conclusion

The integration of AI into science learning materials represents a paradigm shift in science education, redefining not only what students learn but also how they engage with scientific practice and construct scientific knowledge. This chapter has synthesized emerging developments and empirical examples to illustrate how AI is fundamentally transforming the design, representation, and accessibility of science learning materials across multiple dimensions. From embedding authentic scientific practice and enabling adaptive learning experiences to generating multimodal content and expanding accessibility through language and cognitive support, AI technologies offer unprecedented opportunities to enrich and personalize science education. Moreover, the co-creative role of educators and intelligent systems signals a new era of instructional material development, marked by responsiveness, cultural relevance, and iterative design.

These transformations, however, present challenges. The affordances of AI must be balanced with ethical and pedagogical imperatives that ensure transparency, inclusivity, and scientific integrity. Without thoughtful integration, AI-generated materials risk perpetuating epistemic bias (i.e., privileging certain ways of knowing, reasoning, or representing science while marginalizing others), diminishing learner agency, or undermining the pedagogical coherence of science instruction. As such, the development and deployment of AI-supported learning materials must be guided by robust frameworks that center human judgment, critical inquiry, and equitable access.

The implications for science education are profound. The shifting nature of science, as increasingly computational, data-driven, and AI-supported, necessitates corresponding shifts in how science is taught and learned. Science educators, curriculum designers, and policymakers must collaborate to engage with the opportunities and risks posed by AI, creating learning environments that are both innovative and ethically grounded. Professional development efforts should prioritize building educators' capacity to evaluate, adapt, and design AI-supported materials. In parallel, students must be equipped not only to use AI tools but also to critically interrogate their outputs and understand their epistemic boundaries.

In sum, the future of science education in the AI era hinges on the field's ability to leverage AI's transformative potential while upholding the foundational goals of scientific literacy, critical thinking, and social responsibility. AI should not replace human-centered teaching and learning but rather serve as a catalyst for reimagining science education in ways that are more authentic, inclusive, and responsive to the complexities of the contemporary world.

## Acknowledgement

*This material is based upon work supported by the National Science Foundation under Grant No. 2332964 (PI Zhai). Any opinions, findings, and conclusions or recommendations expressed in this material are those of the author(s) and do not necessarily reflect the views of the National Science Foundation.*

## Declare of AI Uses

Parts of this manuscript were prepared with the assistance of generative AI tools. The AI tools were used for language editing or paragraphing of the authors' writing. The authors reviewed, revised, and verified all AI-assisted content for accuracy, appropriateness, and originality. No confidential or sensitive data were entered into AI systems. The final responsibility for the content rests solely with the authors.